\documentclass{elsart3} \usepackage{epsfig}

\begin{document}


\begin{frontmatter}
\title{Sub-10ps Monolithic and Low-power Photodetector Readout}
\author[UH]{L. L.~Ruckman} and  
\author[UH]{G. S.~Varner\corauthref{varner}}
\corauth[varner]{Corresponding author. Tel.: +001 808-956-2987.}
\ead{varner@phys.hawaii.edu}

\address[UH]{Department of Physics and Astronomy, University of Hawaii,
  2505 Correa Road, Honolulu HI 96822, USA}


\begin{abstract}
Recent advances in photon detectors have resulted in high-density
imaging arrays that offer many performance and cost advantages.  In
particular, the excellent transit time spread of certain devices show
promise to provide tangible benefits in applications such as Positron
Emission Tomography (PET).  Meanwhile, high-density, high-performance
readout techniques have not kept on pace for exploiting these developments.  

Photodetector readout for next generation high event rate particle
identification and time-resolved PET requires a highly-integrated,
low-power, and cost-effective readout technique.  We propose fast
waveform sampling as a method that meets these criteria and
demonstrate that sub-10ps resolution can be obtained for an existing
device.

\end{abstract}


\end{frontmatter}


\section{Background}
\begin{figure}[ht]
\vspace*{0mm}
\centerline{\psfig{file=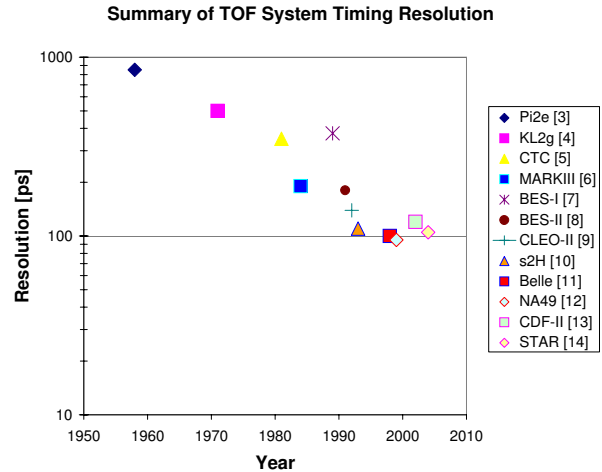,width=3.1in}}
\vspace*{0mm}
\caption{A historical comparison of large Time-Of-Flight system timing
performance, where progress has stagnated.}
\label{TOF_compare}
\end{figure}

Recent developments in high-density, high precision timing
photodetectors are finding applications in Cherenkov detection
techniques for particle identification, as well as medical imaging
applications.  This is true for both traditional
vacuum-based~\cite{PMTs} (Micro-Channel Plate, Hybrid Photo-Diode) and
solid state~\cite{MPPCs} (Geiger-mode APD) photodetectors, and have
resulted in lower cost, higher quantum efficiency, and improved
transit-time-spread (TTS) photon detection options.
Fig.~\ref{TOF_compare} illustrates that without these improvements,
the performance of large-scale Time-Of-Flight (TOF) systems for precision
spectrometers has stagnated at a resolution of about 100ps.

At the same time, the ability for electronics to resolve at the 25ps \space 
or better level has been available for decades, as reviewed in
\cite{porat} and summarized in Table~\ref{psMeasTech}.

\begin{table}[hbt]
\caption{\it An abridged summary, taken from \cite{porat}, of
precision time recording techniques that have been available for
decades.}
\label{psMeasTech}
 \begin{center}
    \begin{tabular}{|l|c|c|} \hline
      {\it Technique }& {\it Resolution [ps] } & {Citation}  \\ \hline\hline
      TDC counter  & 200 (20) & \cite{meyer} (\cite{HPTDC}) \\ \hline
      Vernier (Chronotron) &  $ \sim 15$ & \cite{venable} \\ \hline
      Overlap Coin.(TAC) &  $ \geq 5 $ & \cite{weisberg} \\ \hline
      Start-Stop (TAC) & $ \geq 5 $ & \cite{brafman} \\ \hline
      Microwave & $ \leq 1$ & \cite{guiragossian}  \\ \hline \hline
     \end{tabular}
  \end{center}
\end{table}   

\begin{figure}[ht]
\vspace*{0mm}
\centerline{\psfig{file=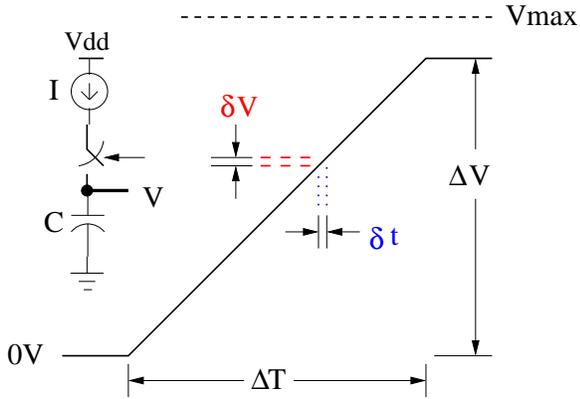,width=3.0in}}
\vspace*{0mm}
\caption{Simple Time-to-Amplitude Converter circuit.}
\label{simp_TAC}
\end{figure}

For instance, consider a very simple Time to Amplitude Converter (TAC)
circuit, as illustrated schematically in Fig.~\ref{simp_TAC}.  When
combined with a modest resolution ADC measurement, this is just about
the simplest possible Time-to-Digital Converter (TDC). In its most
reduced form the circuit consists simply of a current source, a
switch, and a capacitor, as indicated in the figure.

At the start of some event of interest, the switch is closed and
current starts to flow into the capacitor, which is initially at 0V.
Because

\begin{equation}
\Delta Q = C \cdot \Delta V
\end{equation}

and for constant current I, which is $\Delta Q\over \Delta t$, one obtains 

\begin{equation}
\Delta V = {I \over C} \cdot \Delta T.
\end{equation}

After the time interval $\Delta T$, the switch opens and the voltage
stored on the capacitor is a linear function of the ratio of the
current $I$ over the capacitor $C$.  Note that a maximum period $Tmax$
can be measured for a maximum voltage $Vmax$.  Rewriting this
expression we determine the time interval $\Delta T$ simply by
measuring $\Delta V$ and having previously determined the current and
capacitance values (or determining their ratio from calibration).

\begin{equation}
\Delta T = {C \over I} \cdot \Delta V
\end{equation}

Putting in numbers corresponding to typical values for TOF
applications and standard electronics, as listed in
Table~\ref{TOFTACckts}, one sees that  

\begin{equation}
\delta V = {Vmax\over 2^{12}} = 1\; {\rm mV}.
\end{equation}

\begin{table}[hbt]
\caption{\it Typical values for TOF TAC circuits using common electronics components.}
\label{TOFTACckts}
 \begin{center}
    \begin{tabular}{|l|c|} \hline
      {\it Item }& {\it Value }   \\ \hline\hline
      $Vmax$  & 4.095 V   \\ \hline
      Capacitor  & 10 pF   \\ \hline
      Ramp Current & 409.5 $\mu$A \\ \hline
      Full scale &  100 ns (typ.) \\ \hline
      ADC Resolution & 12-bits (4096 counts) \\ \hline
      Noise level & 1mV  \\ \hline \hline
     \end{tabular}
  \end{center}
\end{table}   

This is conveniently at about the amplitude often observed for
board-level and other system pick-up noise in large systems.
Therefore, the least count of this TAC is approximately

\begin{equation}
\delta t = {100\; {\rm ns} \over 2^{12}} = 25\; {\rm ps}
\end{equation}

and for an ideal binary interpolation of this least count, one could
do even a factor of $1\over \sqrt{12}$ better.  Non-linearites due to
imperfections and temperature dependence in the current source and
switch can degrade performance, but have been addressed in a number of
commercial modules.  For high rate environments, variants of this same
basic scheme involving multiple ramps, have been employed for fast TOF
measurement in an online trigger~\cite{orlov}, as well as hit logging
prior to trigger formation~\cite{BelleTS}.

This performance is so much better than that for the $\sim 100$ps
limit mentioned earlier that even if the TDC contribution were made zero, the
overall timing would only improve by 3ps:

\begin{equation}
\sigma = \sqrt{ (100\; {\rm ps})^2 - (25\; {\rm ps})^2 } \sim 97 \; {\rm ps}
\end{equation}

Therefore, the culprit must reside in the method of conveying the
leading edge timing of the photodetector pulse into a signal that
these proven methods can use.


\section{Amplitude Dependent Effects}

\begin{figure}[ht]
\vspace*{0mm}
\centerline{\psfig{file=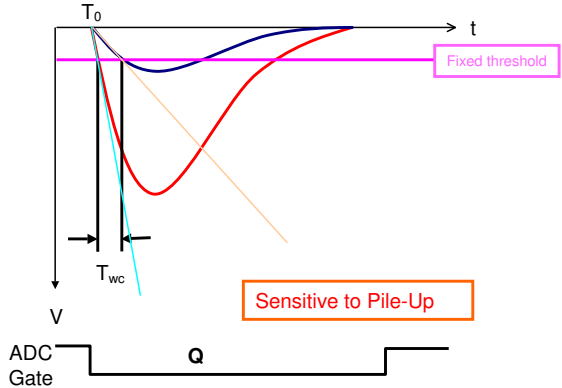,width=3.0in}}
\vspace*{0mm}
\caption{Example of the so-called time walk effect and its correction
for the Belle TOF case.}
\label{TimeWalk}
\end{figure}

This well-known issue of ``time walk'' is illustrated in
Fig.~\ref{TimeWalk}, where a photodetector output signal shape is
amplitude dependent.  Time walk corrects are used to 
compensate for the time of a threshold
crossing slews to later times for smaller pulses.

\begin{figure}[ht]
\vspace*{0mm}
\centerline{\psfig{file=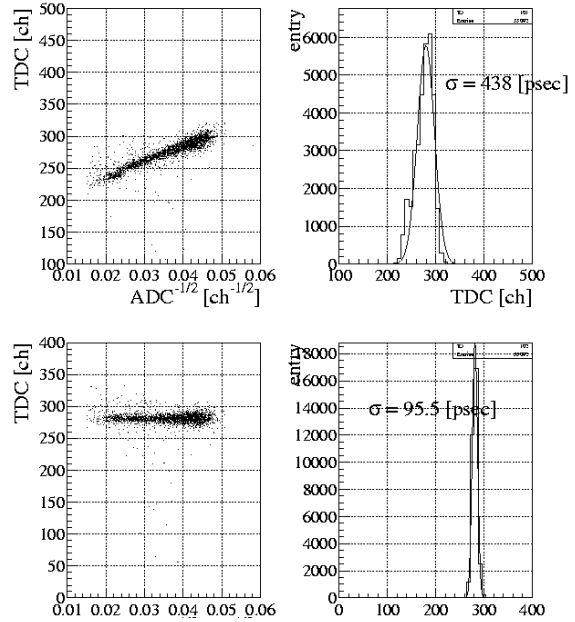,width=3.0in}}
\vspace*{0mm}
\caption{Amplitude dependence and result after applying TWC.}
\label{TWCeffect}
\end{figure}

For detector signals with large amplitude variations, the effect can
be quite sizeable and must be corrected.  A simple and proven method
is to integrate the charge of the pulse within a gate around the
signal of interest.  While the leading edge of the signal can be
complex, and can be modelled by a power-law, the observable of
interest is the dependence on the pulse-integrated charge.  This is
illustrated in Fig.~\ref{TWCeffect}, where the dependence observed for
the Belle TOF system~\cite{Belle} is consistent with many previous
observations as being proportional to approximately $1\over \sqrt{Q}$.

Plotted as a function of $ADC^{-{1\over 2}}$ in the upper left graph
of Fig.~\ref{TWCeffect}, this dependence appears then as a linear
expression.  The effect of applying the correction is shown in the
lower plots, where the amplude dependence has been removed and more
than a factor of 4 in timing resolution improvement is observed.
However, to further reduce scatter in the distribution requires a different
or better parameterization. 

For signals with smaller amplitude variations, such as in single
photon detection, Constant Fraction Discrimination (CFD) techniques can be
used instead, and may be combined with multi-hit
devices~\cite{HPTDC,MTS} to achieve good timing performance.  However,
pile-up and rate effects need to be studied for each application.
Moreover, all of the difficult work in making this circuitry
functional is pushed onto the discriminator circuit.  To reduce the
effect of overdrive, the dependence of comparator output time upon
amplitude above threshold, often a large amount of power may be
required, making the circuitry difficult to densely integrate.  Due to
the intrinsic memory of the CFD technique, it is susceptible to pile-up effects.

A number of techniques, such as multi-level thresholding, have been
proposed to directly measure the risetime and provide a compensating
correction.  Each threshold measurement may require an additional
amplitude dependent correction to obtain the best possible timing.
Adding more and more thresholding levels, one asymptotically
approaches the penultimate technique, which is waveform sampling.


\section{Waveform Sampling}

A measurement complication at high event rates is pile-up, where the
timing signal of interest may be superimposed on the tail of a
preceding pulse.  Without correction, and often being unaware that
this superposition is present in the data, an incorrect timing
determination will be made.  If affordable and fast
enough~\cite{BLAB1}, and if the aperture jitter is small enough, waveform
sampling provides maximum information extraction.  The benefits may be
summarized as follows:

\vspace{0.1in}

\begin{itemize}
\item Pile-up impact extraction
\item Fit to actual pulse shape
\item Average to reduce single measurement errors
\item Observe changes to pulse shape over time
\end{itemize}

\vspace{0.1in}

where the last issue can be particularly important for detectors that
undergo radiation damage or aging effects such that the intrinsic
pulse shape evolves with time.


\section{The BLAB1 Monolithic Sampler}

To demonstrate the viability of the waveform sampling technique, we use the first
Buffered LABRADOR (BLAB1) ASIC~\cite{BLAB1}.  This is a $2^{16}$ (64k)
cell deep waveform sampling ASIC that was developed as a deep-sampling
extension of the successful LABRADOR ASIC~\cite{LAB3}.  The key
parameters of this device are summarized in Table~\ref{BLAB1_specs}.

\begin{table}[hbt]
\caption{\it Specifications of the BLAB1 ASIC used as a waveform
sampling device to demonstrate the precision timing extraction technique.}
\label{BLAB1_specs}
 \begin{center}
    \begin{tabular}{|l|c|} \hline
      {\it Item }& {\it Value }  \\ \hline\hline
      Sampling Input Channels  & 1  \\ \hline
      Storage rows &  128 \\ \hline
      Storage cells/row &  512 \\ \hline
      Total storage cells &  65,536 \\ \hline
      Sampling speed (GSa/s)&  0.1 - 6.0  \\ \hline
      Storage record &  10.9 - 655 $\mu$s  \\ \hline
      Operation mode & continuous storage/readout \\ \hline \hline
     \end{tabular}
  \end{center}
\end{table}   

\begin{figure}[ht]
\vspace*{0mm}
\centerline{\psfig{file=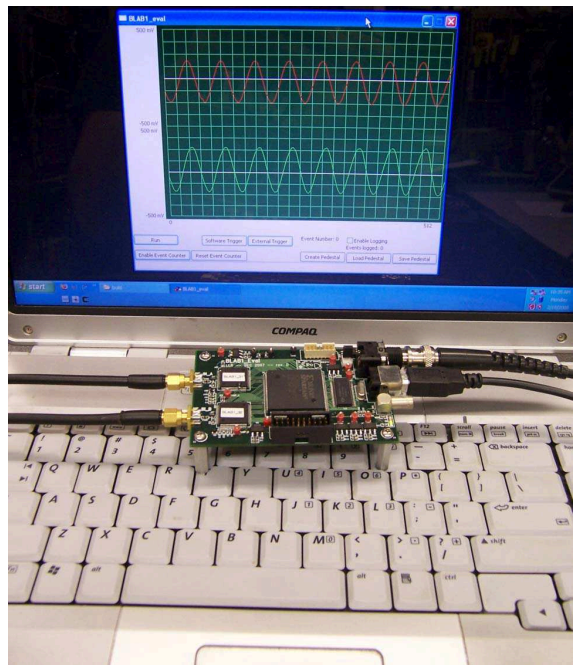,width=3.0in}}
\vspace*{0mm}
\caption{Photograph of the ``oscilloscope on a chip'' test set-up,
where the 2 BLAB1 ASICs are at the left, near the SMA input
connectors.}
\label{TestSetUp}
\end{figure}

The evaluation system consists of the small circuitboard, containing 2
BLAB1 ASICs, and read out over USB2.0 by a laptop running a custom
data acquisition program based on the wxWidgets tool kit.  A photo of
this compact measurement system is displayed in Fig.~\ref{TestSetUp}.
This compact configuration can turn any PC (or laptop) into a
high-performance digital signal oscilloscope.

\section{Precision Timing Performance}

All of the standard precision timing measurement techniques discussed
at the outset suffer from practical limitations in actual application,
which has served to degrade the realized large system performance.  In
the end, one simply cannot do better than having a high-fidelity
``oscilloscope on a chip'' for every sensor channel.  Cost and data
volume precluded this type of waveform recording until recent
generations of SCA ASICs \cite{LAB3,ATWD,DRS,Stefan,SAM} demonstrated
such techniques were practical, especially for large systems.

\subsection{Calibration}

\begin{figure}[ht]
\vspace*{0mm}
\centerline{\psfig{file=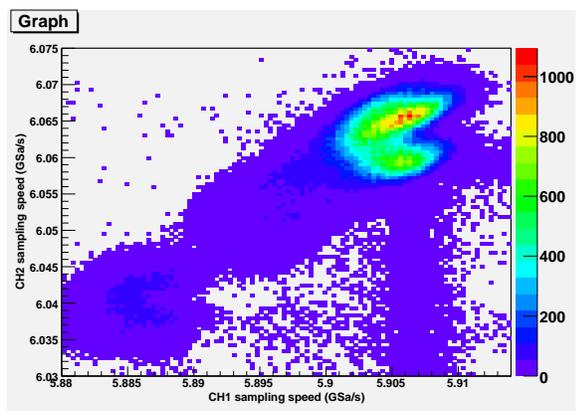,width=3.0in}}
\vspace*{0mm}
\caption{Event-by-event sampling speed comparisons of two BLAB1
ASICs indicating a common temperature dependence term.  The sampling speed
is continually monitored and used in calibration to achieve the
maximum timing performance.}
\label{SS_CORREL}
\end{figure}

In order to achieve fine timing resolution, a number of calibrations
need to be performed and corrections need to be applied.  As reported
in earlier devices, the temperature dependence~\cite{BLAB1,LAB3} of
the sampling baseline must be compensated.  The effect of this
$0.2\%/^{\circ}C$ temperature dependence is seen in
Fig.~\ref{SS_CORREL}.

In order to address bin-by-bin timing width differences, a couple of
different calibration techniques have been tried.  The first utilizes
a sine wave zero-crossing technique used for calibrating the LAB3
ASIC\cite{LAB3}.  That technique works best when the frequency of the
sine wave is such that the measured interval between zero crossings
can be uniquely assigned to a limited number of bins between
successive crossings. 

\begin{figure}[ht]
\vspace*{0mm}
\centerline{\psfig{file=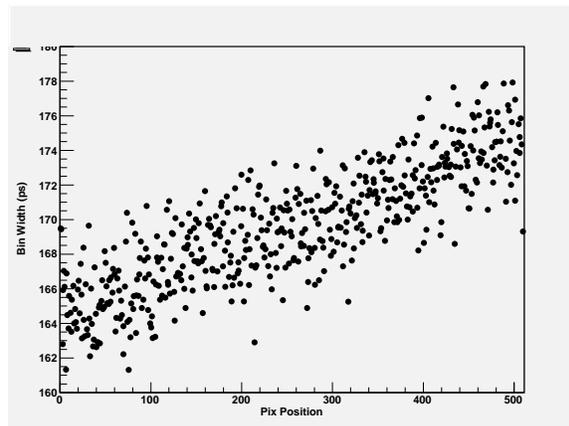,width=3.0in}}
\vspace*{0mm}
\caption{Aperture width determination for individual sample cells via the
histogram occupancy technique described in the text.}
\label{Occup}
\end{figure}

Due to sine wave curvature, this technique has an irreducible
systematic error that is a function of sample rate.  A more successful
technique is to histogram the zero crossings of a sine wave and use
the bin occupancy to derive the effective aperture width, as may be
seen in Fig.~\ref{Occup}.

When a sampling strobe is inserted into the BLAB1, the strobe
propagates across a Switched Capacitor Array (SCA) row and exits the
ASIC via a monitor pin.  The sampling speed is determined by measured
this propagation delay, a method implemented by creating a feedback
loop with the insertion-extraction chain and an inverter located
inside the companion FPGA.  This feedback inverter is only connected
after waveform readout of a triggered event and then disconnected once
the sampling speed has been measured for that particular triggered
event.
%
A so-called Ripple Carry Out (RCO) period may then be expressed as
\begin{equation}
T_{\rm total} = 2*(T_{\rm latency} + T_{\rm waveform})
\end{equation}
where the factor of 2 comes from the FGPA feedback inverter toggling
the RCO period, $T_{\rm waveform}$ is the SCA propagation time, and
$T_{\rm latency}$ is the additional time accrued due to internal FPGA
routing, circuit board routing, and routing to and from the SCA within
the ASIC.  To measure $T_{\rm total}$ an FPGA-based Time-to-Digital
Converter (TDC) is used~\cite{JINST}.  Both phases of a 250 MHz clock
are used by the TDC to obtain a least count of 2ns.  Improved
resolution is obtained by integrating the RCO period for 2048 cycles,
corresponding to an aggragate single-cycle least count of 0.977 ps.
$T_{\rm latency}$ is obtained by subtracting $T_{\rm waveform}$ from
$T_{\rm total}$, where $T_{\rm waveform}$ is determined from a fit to
a fixed frequency sinusoidal signal.  The sampling speed is then
\begin{equation}
f_{\rm SPD} = (T_{\rm waveform}/{\rm Row}_{\rm length})^{-1}
\end{equation}
where $f_{\rm SPD}$ is the sampling speed and ${\rm Row}_{\rm length}$ is
the number of SCA cells in a row, which for the BLAB1 ASIC is 512
cells.  As reported in Ref.~\cite{BLAB1}, the sampling timebase is
temperature dependent.  This effect may be seen as the correlated
portion of Fig.~\ref{SS_CORREL}.  By measuring the BLAB1 timebase,
these timebase shifts can be monitored and corrected offline.

\subsection{Ideal Test Signals}

\begin{figure}[ht]
\vspace*{0mm}
\centerline{\psfig{file=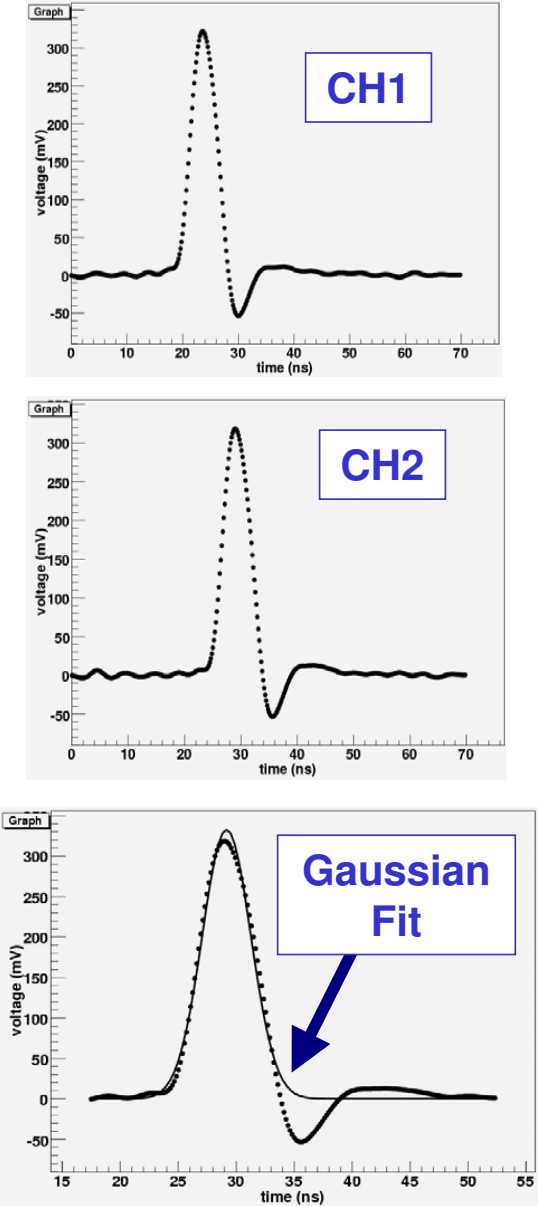,width=3.0in}}
\vspace*{0mm}
\caption{Pulses recorded in two separate BLAB1 devices, with a small
timing offset.  A gaussian fit, as shown in the bottom waveform, is
used to extract the event-by-event time of each signal.}
\label{gaus_shape}
\end{figure}

An idealized test pulse of fixed amplitude and width was generated by
taking an Agilent 33250A arbitrary waveform generator pulse output, AC
coupling the signal with an inline capacitor, RF splitting the signal,
and waveform sampling both signals using two separate BLAB1 ASICs with
additional cable delay for the second channel.  This AC coupling
served as a filter and shaped the pulse into an almost gaussian shape,
as seen in Fig.~\ref{gaus_shape}.

\begin{figure}[ht]
\vspace*{0mm}
\centerline{\psfig{file=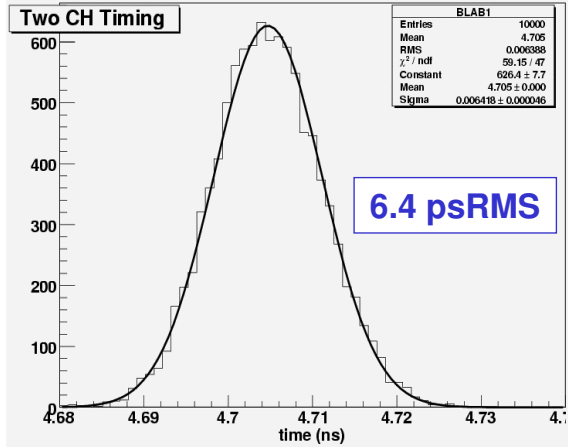,width=3.0in}}
\vspace*{0mm}
\caption{Time difference between a pair of idealized pulses.  Each
edge can be inferred to be extracted a factor of $\sqrt{2}$ better, or
about 4.5ps RMS.}
\label{best_ps}
\end{figure}

The result obtained for taking the time difference of a simple
gaussian fit to the signals in both channels is plotted in
Fig.~\ref{best_ps}.

\subsection{Varying Sampling Window Position}

\begin{figure}[ht]
\vspace*{0mm}
\centerline{\psfig{file=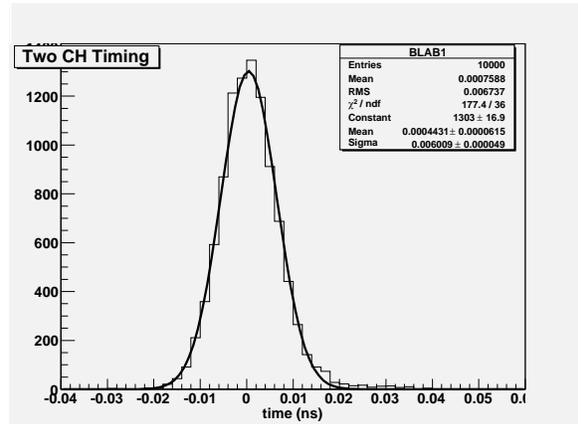,width=3.0in}}
\vspace*{0mm}
\caption{Measured time resolution by taking the cross correlation of
two ideal pulse waveforms.}
\label{10psRes}
\end{figure}

The two pulses were then recorded without regard to the sampling
strobe phase.  This allows the two pulses to move around within the
waveform sampling window.  During the analysis, a bin-by-bin timebase
correction and an event-by-event sampling speed correction were
applied.  The time was measured by taking the cross correlation of the
two waveforms.  The cross correlation was done by taking the Fast
Fourier Transform (FFT) of both waveforms, complex multiplying one FFT
by the complex conjugate of the other FTT, and Inverse Fast Fourier
Transforming (IFFT) this product to get the cross correlation waveform
in the time domain.  The time position of the maximum point in the
cross correlation waveform is the time difference between the best
estimator of the location of the two pulses.  Because the sampling
speed between the two ASIC chips vary from each other and the time
spacing between the bins is not constant, before FFT a fifth order
data spline is applied at a fixed time spacing very close to the
average time interval.  This is done to keep the sampling spacing
between points constant and minimize frequency domain distortion.  To
get finer time spacing in the cross correlation waveform without
distortion, we symmetrically insert zeros in the cross correlation FFT
for frequencies greater than the Nyquist frequency and less than the
negative Nyquist frequency before the IFFT operation.  The time
resolution obtained using this technique is shown in
Fig.~\ref{10psRes}, corresponding to an RMS of 6.7 RMS or a $\sigma $
of 6.0 ps.

For complex curvature along the leading edge of the signal, the timing
resolution obtained is seen to be rather sensitive to the method
choosen to characterize the signal ``hit'' time.  Unless the
photodetector signal is for a single p.e. quanta, the actual shape can
be rather complex and dependent upon photon arrival statistics.  Even
in this simple case, noise and aperture systematics upon the leading
edge can be important and can also be reduced by using multiple
samples to fit to an analytic signal shape.  In general, the estimate
error can improve as something like 1/$\sqrt{N}$ for N samples along
the leading edge.  This is perhaps the most powerful aspect of having
the full waveform samples to fit.  Individual sampling errors can be
averaged out.  Examples are provided in the following subsection,
where it is clear that at the sampling rates being studied, this
waveform recording technique logs many samples on the leading edge,
which can be used to improve the signal timing extraction.

\subsection{PMT signal observation}

\begin{figure}[ht]
\vspace*{0mm}
\centerline{\psfig{file=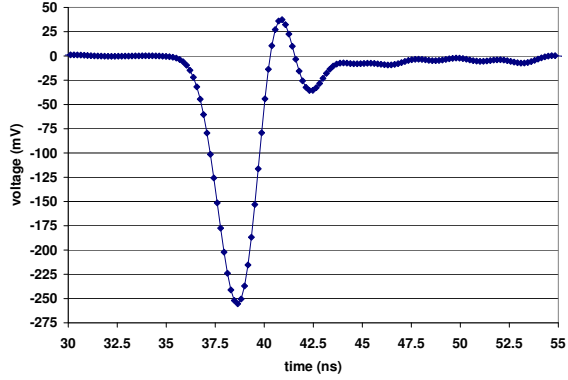,width=3.0in}}
\vspace*{0mm}
\caption{Example waveform used in this analysis as recorded from a
Burle 85011 Micro-Channel Plate PMT.}
\label{PMT_plot}
\end{figure}

A convenient feature of the BLAB1 ASIC is that a PMT output
transmitted over a 50$\Omega$ coaxial cable can be directly connected
to the BLAB1 input.  A waveform recorded from a Burle 85011
Micro-Channel Plate PMT is displayed in Fig.~\ref{PMT_plot}. 

This near ideal waveform is utilized as the simulation template in the
following section on performance limits from simulation.

\subsection{MCP-PMT signal timing}

\begin{figure}[ht]
\vspace*{0mm}
\centerline{\psfig{file=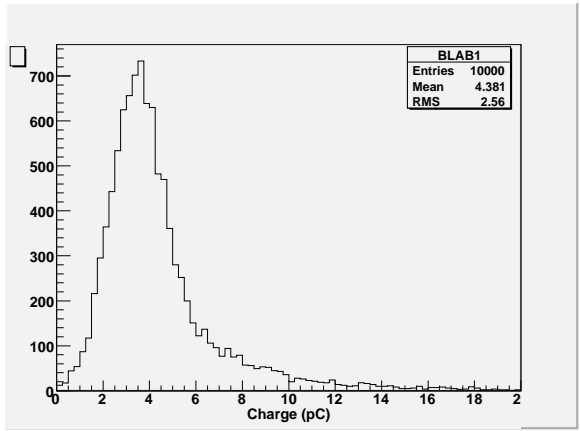,width=3.0in}}
\vspace*{0mm}
\caption{Pulse height distribution for the signals used in this timing analysis.
}
\label{MCP_GAIN}
\end{figure}

A set of 10k triggered (cosmic and dark count) events were recorded,
as illustrated for an example signal in Fig.~\ref{PMT_plot}.  The
signal amplitude was allowed to fluctuate and the measured distribution
of integrated charge is shown in Fig.~\ref{MCP_GAIN}.

\begin{figure}[ht]
\vspace*{0mm}
\centerline{\psfig{file=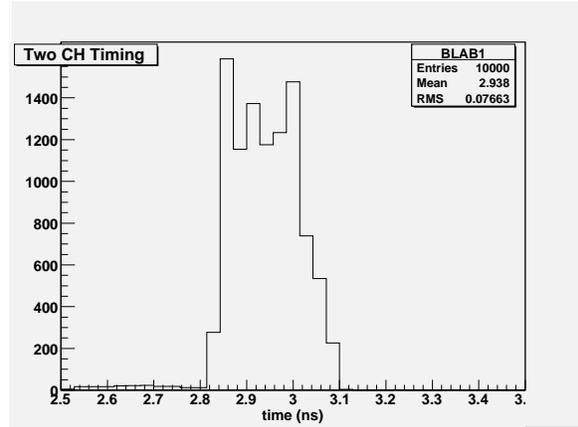,width=3.0in}}
\vspace*{0mm}
\caption{Raw timing performance after simple timebase correction.}
\label{Burle_before}
\end{figure}

Ideally one would use a fast timing (ps) laser to do this measurement,
however the Transit Time Spread of the photodetector would then be an
issue.  Instead we use the same pulse, split it, and record the signal
on two storage channels.  After applying timebase and time bin
corrections, the raw timing difference distribution seen in
Fig.~\ref{Burle_before} is obtained.

\begin{figure}[ht]
\vspace*{0mm}
\centerline{\psfig{file=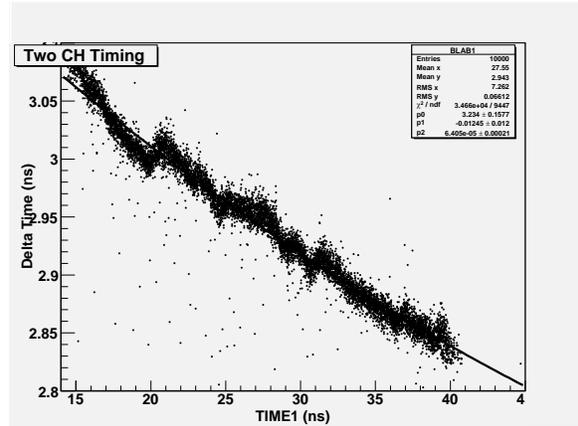,width=3.0in}}
\vspace*{0mm}
\caption{Observed residual timing structure, indicating a limitation
in the previous calibration procedure.}
\label{MCP_BURLE_STRUCTURE}
\end{figure}

This distribution is rather wide and is related to a residual timing
miscalibration.  This can be seen by looking at the time difference as
a function of the peak time determined on the first channel, as
shown in Fig.~\ref{MCP_BURLE_STRUCTURE}.

\begin{figure}[ht]
\vspace*{0mm}
\centerline{\psfig{file=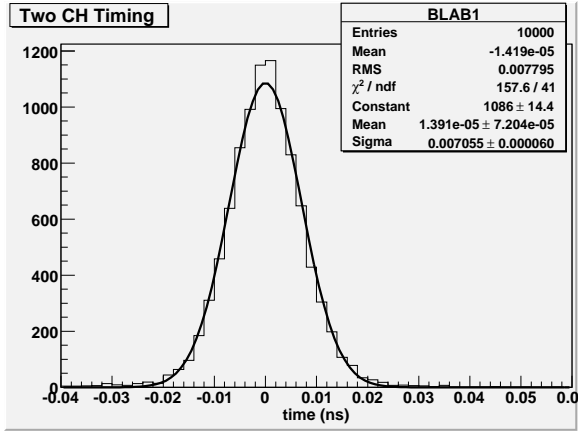,width=3.0in}}
\vspace*{0mm}
\caption{Time distribution after applying an optimized, binned timing correction.}
\label{MCP_BURLE_7p8PSRMS}
\end{figure}

We created a secondary bin-by-bin timing correction related to the time 
position of the first peak using the data from Fig.~\ref{MCP_BURLE_STRUCTURE}.  
After applying this secondary tinming correction, Fig.~\ref{MCP_BURLE_7p8PSRMS}
illustrates the improvement in timing performance.  


\begin{figure}[ht]
\vspace*{0mm}
\centerline{\psfig{file=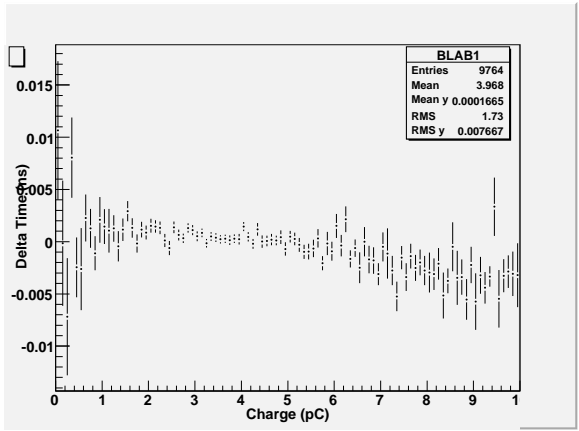,width=3.0in}}
\vspace*{0mm}
\caption{Improved time difference residual as a function of integrated
pulse charge.}
\label{MCP_BURLE_Q_AFTER}
\end{figure}

For this measurement, the amplitude of the MCP signal varies from
event to event.  For the waveform data, we calculate the charge of the
signal by summing all the waveform data points, multiplying by the
average timebase, and dividing by the termination resisitor.  After
applying all the corrections previously mentioned, a profile histogram
of the charge versus time in Fig.~\ref{MCP_BURLE_7p8PSRMS} is shown in
Fig.~\ref{MCP_BURLE_Q_AFTER}, which illustrates that only a slight
amount of residual time walk is remains after applying the correction.

\section{Limitations}
While the analog bandwidth of the BLAB1 is adequate for many RF
recording applications, a higher bandwidth device will be explored,
based upon the lessons learned from this first device.  In particular,
the fanout structure and design of the analog amplifier tree is being
scrutinized and improved in simulation.  It is hoped that an almost
arbitrarily large storage depth can be accommodated up to 1GHz of
analog bandwidth through a careful layout of the buffer amplifier
cascade array.  

To gain insight into the parameters that are most important to
attaining the highest possible timing precision, a series of
simulation studies were performed.  In each run a series of 10k events
were generated with a time offset equal to that used in the previous
section.  A real pulse template shown in Fig.~\ref{PMT_plot} has
been used, and is perturbed as described for each of the following
cases.  The simulation outputs the distorted waveforms into the same
data structure as used by the acquisition software and the same
cross-correlation method described earlier is applied.

\subsection{Random Noise}

\begin{figure}[ht]
\vspace*{0mm}
\centerline{\psfig{file=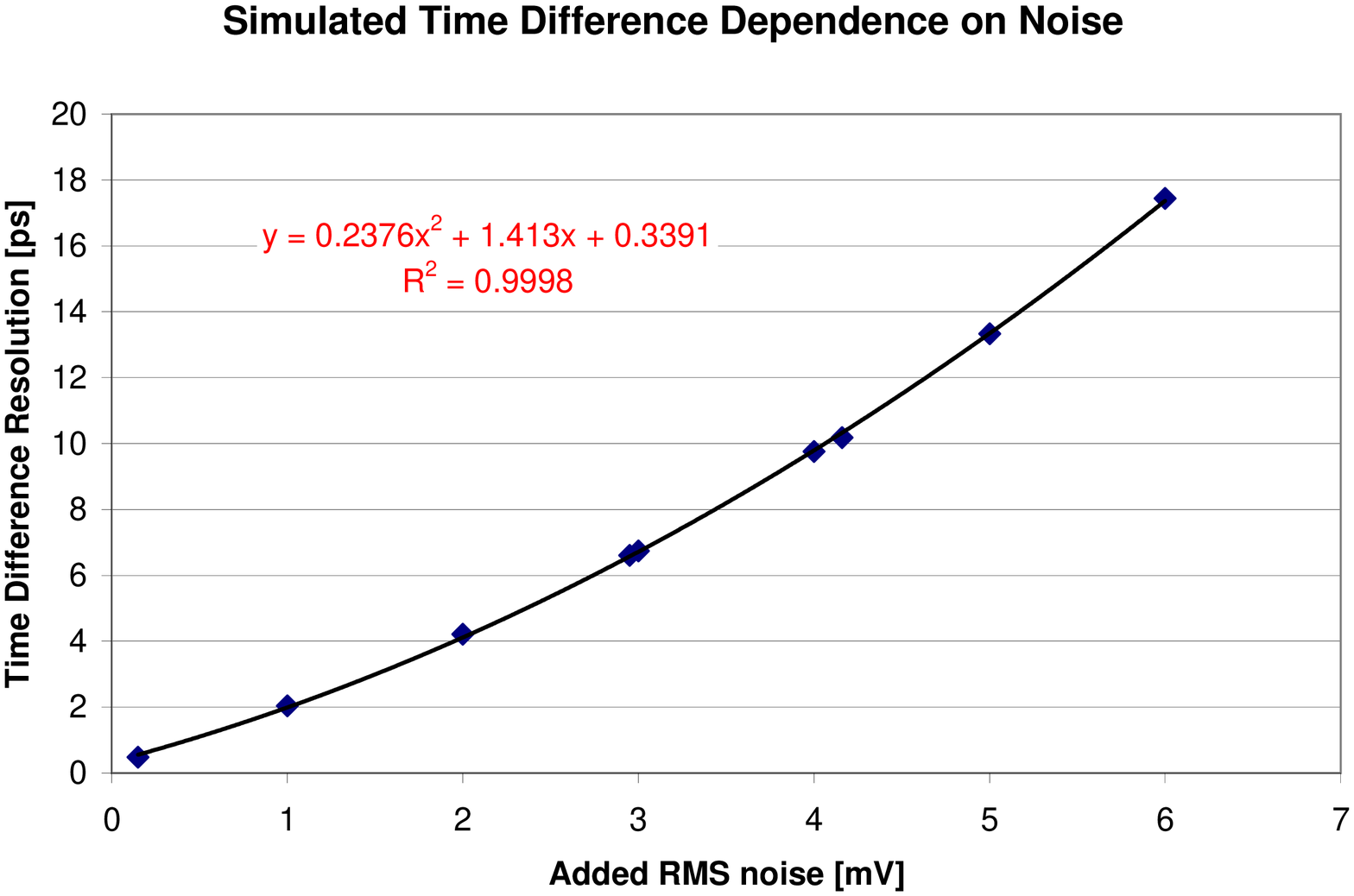,width=3.0in}}
\vspace*{0mm}
\caption{Fitted time resolution as a function of added noise.}
\label{Sim_noise}
\end{figure}

The result due to the random addition of noise to each sample of the
two waveforms is illustrated in Fig.~\ref{Sim_noise}.  Comparing the
curve with the measured result, an effective noise of $\sim 3{\rm mV}$ is
indicated.

However, since we know that the noise in each sample is only about
1mV, additional contributions need to be considered.

\subsection{Sample Aperture Jitter}

\begin{figure}[ht]
\vspace*{0mm}
\centerline{\psfig{file=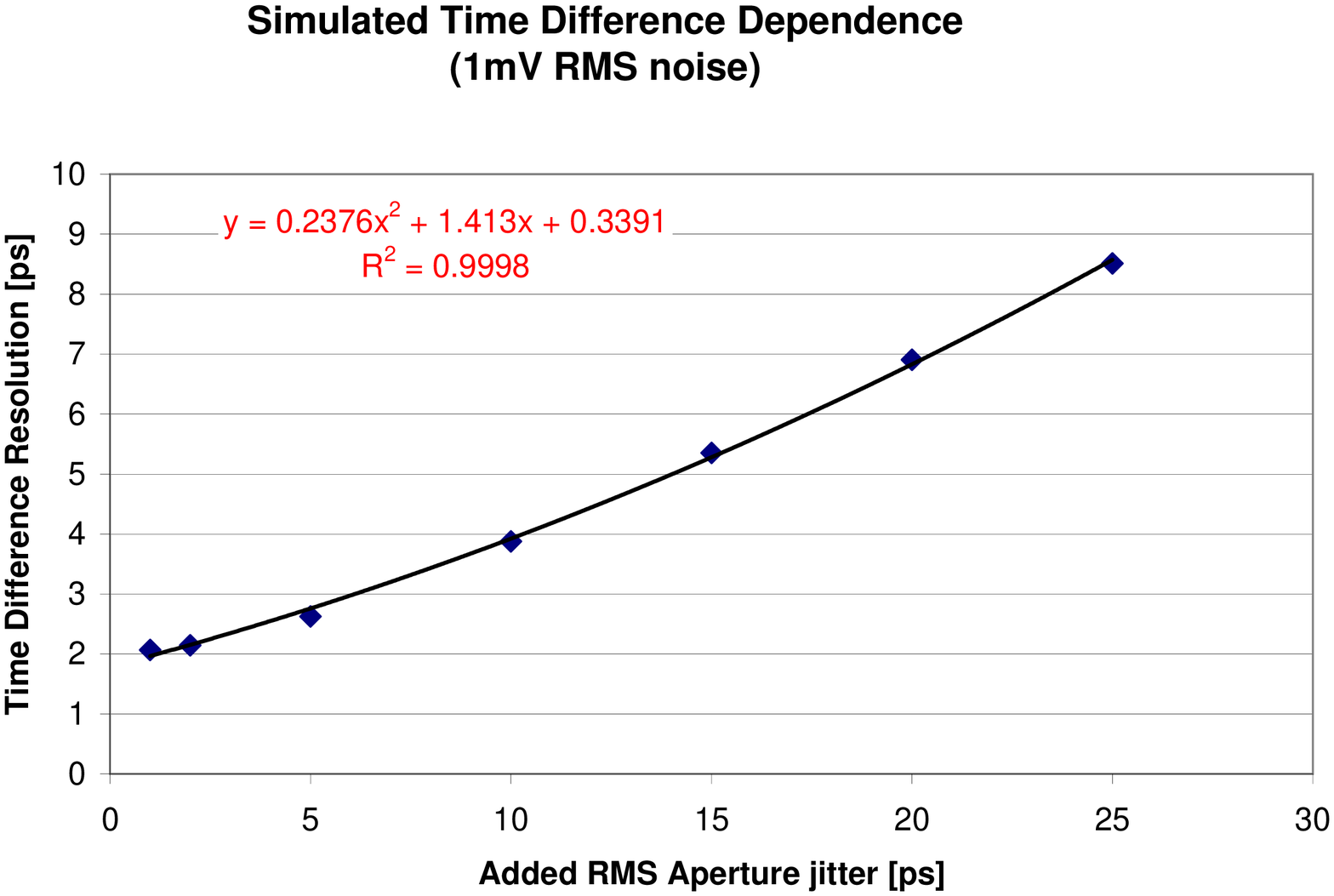,width=3.0in}}
\vspace*{0mm}
\caption{Timing resolution with 1mV noise as a function of aperture jitter.}
\label{Samp_jitter}
\end{figure}

Another expected contribution to the extraction of the timing
resolution is jitter on the timing samples.  Using the same simulation
configuration, the sample noise is now fixed and the effect of adding
a randomized aperture jitter is plotted in Fig.~\ref{Samp_jitter}.

\begin{figure}[ht]
\vspace*{0mm}
\centerline{\psfig{file=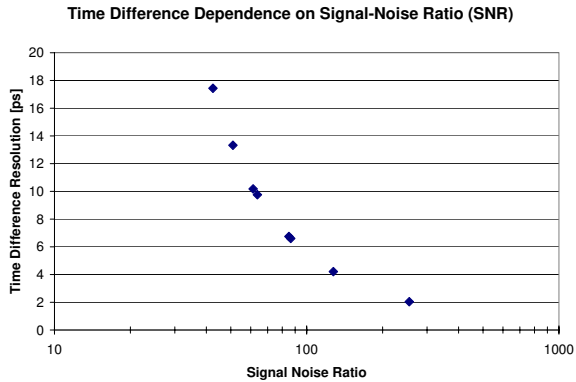,width=3.0in}}
\vspace*{0mm}
\caption{Impact of effective sampling noise on timing resolution,
expressed in SNR.}
\label{SNR_extra}
\end{figure}

The local signal slope is used to determine a voltage offset expected
by sampling at the wrong time.  Since the sampling period is small
compared with the typical voltage step between samples, a rather large
jitter is needed to reproduce the observed time resolution.  From
direct measurements of timebase jitter, we expect a far smaller value
than is required to explain the time resolution observed.  It is noted
that the two effects are degenerate in the sense that they lead to a
sample voltage dispersion from the ideal value.  Considering this, one
can express the time difference resolution in terms of a combined
value characterized in terms of a Signal to Noise Ratio (SNR).  This
relationship is seen in Fig.~\ref{SNR_extra}, where it is clearly seen
that an effective SNR of 100 or greater is desired to obtain time
resolutions of 5ps or better.

\subsection{Sampling Timebase Jitter}

\begin{figure}[ht]
\vspace*{0mm}
\centerline{\psfig{file=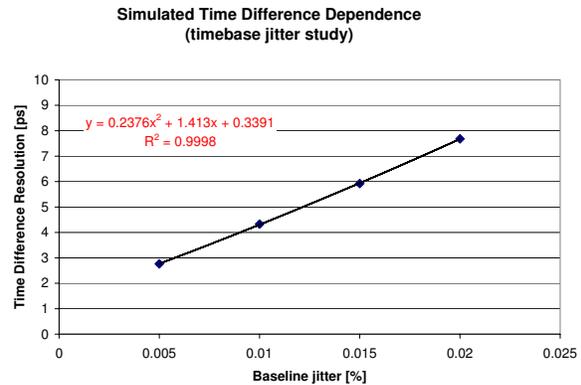,width=3.0in}}
\vspace*{0mm}
\caption{Time resolution with 1mV noise, 1ps of sample jitter, and
as a function of event-by-event uncorrected timebase jitter.}
\label{timebase_jitter}
\end{figure}

While the pedagogical illustrations above are helpful in understanding
what limits the time resolution that can be obtained, there is an
additional contribution that is an artifact of the use of the BLAB1
ASIC.  Specifically, it contains no dedicated Delay Lock Loop (DLL)
circuitry, and thus the sampling timebase is sensitive at the level of
about 0.2\%/$^{\circ}C$~\cite{BLAB1}.  Even though an event-by-event
correction is performed, as described previously, it is quite probable
that a residual correction error contributes to the time resolution.
A simulation is performed as shown in Fig.~\ref{timebase_jitter},
where 1mV of noise is added in quadrature with 1ps of sample aperture
jitter on top of an overall event-by-event sampling timebase jitter.
This jitter is assumed to be uncorrelated between the two ASICs.

In the limit that the known 1mV noise only contributes 2ps to the
overall time resolution and that the sample aperture jitter is
negligible, this effect is seen to be dominant.  Quantitatively, this
corresponds to about a 10\% error on correcting the event-by-event
sampling timebase.  If true, this can be improved through better
timebase stabilization in a next version of the ASIC~\cite{BLAB1}, or
through the use of other devices as described next.

%

\section{Other Implementations}

The waveform sampling technique described is more general than the
particular ASIC presented.  Indeed if price and power are no object, a
number of commercial digitizers are available, with new devices
continually coming to market.  In general those devices, besides being
rather power hungry, also suffer from a limited dynamic range.
However, for many applications, they are already or will become viable.  

In the near term, a number of custom ASICs with better analog
resolution~\cite{delagnes}, faster sampling~\cite{jfg} and faster
readout~\cite{drs4}, will be available to implement these waveform
sampling techniques.  Further studies of the performance limits on
timing signal extraction using this waveform sampling technique is a
high priority for the follow-on BLAB2 ASIC~\cite{BLAB1} as well.
In the longer term, these low-cost devices show great promise to
replace standard ADC and TDC techniques in many future high energy and
particle astrophysics detectors~\cite{PD07}.  As we have demonstrated,
they should be able to do so with excellent timing precision.

%



\section{Summary}

A waveform sampling technique has been shown to obtain sub-10ps timing
resolution for readout of photodetector signals.  Measurements were
made with an existing compact, low-power and pipelined readout ASIC.
Such a technique can be scaled up to very large channel counts in a
cost-effective way.


\section{Acknowledgements}
The authors gratefully acknowledge the generous support of the MOSIS
Educational Program, which provided the fabrication of the BLAB1 ASIC
prototype through their University Research Program.  Testing was
supported in part by Department of Energy Advanced Detector Research
Award \# DE-FG02-06ER41424.  This manuscript is dedicated to Sim
Edward Varner, who passed away during its preparation.


\end{document}